# Effect of Size Distribution, Skewness and Roughness on the Optical Properties of Colloidal Plasmonic Nanoparticles

Rituraj Borah[a,b], Sammy W. Verbruggen[a,b]*

[a]Sustainable Energy, Air & Water Technology (DuEL), Department of Bioscience Engineering,

University of Antwerp, Groenenborgerlaan 171, 2020 Antwerp, Belgium

[b]NANOlab Center of Excellence, University of Antwerp, Groenenborgerlaan 171, 2020 Antwerp,

Belgium

*Sammy.Verbruggen@uantwerpen.be

**Abstract**

It is a generally accepted idea that the particle size distribution strongly affects the optical spectra of colloidal plasmonic nanoparticles. It is often quoted as one of the main reasons while explaining the mismatch between the theoretical and experimental optical spectra of such nanoparticles. In this work, these aspects are critically analyzed by means of a bottom up statistical approach that considers variables such as mean, standard deviation and skewness of the nanoparticle size distribution independently from one another. By assuming normal and log-normal distributions of the particle size, the effect of the statistical parameters on the Mie analytical optical spectra of colloidal nanoparticles was studied. The effect of morphology was also studied numerically in order to understand to what extent it can play a role. It is our finding that the particle polydispersity, skewness and surface morphology in fact only weakly impact the optical spectra. While, the selection of suitable optical constants with regard to the crystallinity of the nanoparticles is a far more influential factor for correctly predicting both the plasmon band position and the plasmon bandwidth in theoretical simulations of the optical spectra. It is shown that the mean particle size can be correctly estimated directly from the plasmon band position, as it is the mean that determines the resonance wavelength. The standard deviation can on the other hand be estimated from the intensity distribution data obtained from dynamic light scattering experiments. The results reported herein clear the ambiguity around particle size distribution and optical response of colloidal plasmonic nanoparticles.

## 1. Introduction

Over the years, plasmonic nanostructures have gained considerable attention in various application areas such as solar energy harvesting[1], photocatalysis[2], sensing[3], photothermal therapy[4][5], *etc*. Thus, well-controlled fabrication of such nanostructures has emerged as an important field of research. For this reason, wet chemical synthesis of plasmonic nanoparticles has gained considerable attention over the years[6][7]. The wet chemical route is advantageous due to its facile procedures, cost-effectiveness and high throughput when compared with clean-room fabrication techniques[8] [9]. The wet-chemical route can even facilitate synthesis of nanoclusters of a few hundreds of atoms or less,[10][11] prompting considerable efforts towards the development of large scale continuous reactor systems for the synthesis of nanoparticles.[12][13] A major challenge in the wet chemical route, however, is the lack of full control over all nanoparticle characteristics. Due to the inherent randomness in the nucleation and the subsequent growth process, the size distribution of the nanoparticles is difficult to control.[14] Polydispersity is also a challenge in the synthesis of nanoparticles by pulser laser abalation.[15][16] Thus, the minimization of the polydispersity is one of the important challenges of this synthesis route as highly polydisperse or irregular particles poses difficulties for further applications.[17][18][19] It is widely believed and accepted that the optical properties of plasmonic nanoparticles strongly depend on the size, polydispersity and non-uniformity of the particle morphology.[20][21][22][23]It has also been shown that polydispersity has a strong effect on the total energy absorption and thus, on the plasmonic heat generation.[24] However, after a detailed analysis the effect of particle polydispersity seems to be rather speculative as no study can be found elucidating the exact role of the size distribution in determining the optical spectral response.

For nanoparticles of 10 nm in size or bigger, the classical electromagnetics theory can be used with sufficient accuracy to calculate the optical properties, as the quantum size effects are not significant enough for larger clusters. This requires an accurate set of optical constants determined experimentally for the given metal or the alloy composition with similar crystal properties. The high optical absorbance obtained at the resonance frequency in the optical spectra of the nanoparticles, resulting from Localized Surface Plasmon Resonance (LSPR), is captured quite realistically by solving Maxwell's equations. While for Au, the agreement between the calculated spectra of single isolated nanoparticles using classical Mie theory, which is the analytical solution of the Maxwell's equations, and the experimental spectra of the colloidal nanoparticles is often satisfactory[25], the same is not at all the case for Ag. Often, the simulated optical spectra are blue shifted by over 10 nm, accompanied by a much smaller bandwidth (characterized by the full width half maximum, FWHM) relative to the experimental spectra.[26] Apart from the general deviation of the theoretical LSPR wavelength and bandwidth from the experimental, the higher order resonance peaks are not visible in the optical spectra of large Ag nanoparticles (>100 nm).[8] Apart from the uncertainty in the experimental optical constants used in the calculations and polydispersity, morphological irregularities are also often stated to explain the large difference between experimental spectra and theoretical optical spectra obtained for perfectly spherical and monodisperse nanoparticles.[27][28] The Ag optical data from two highly cited literature sources, Palik[29] and Johnson and Christy[30], show significant differences in the imaginary part of the refractive index, possibly due to the difference in the sample preparation methods that determine the crystallinity. Generally, such discrepancies are observed across various experimental data sets.[31][32][33][34]Thus, with many possible factors such as particle crystallinity, morphology, particle size distribution, *etc*., to be playing their roles, it is our aim to understand the effects of these factors independently from one another. Further insight into these factors will be helpful in the accurate prediction of optical response of Ag nanostructures in general, which is in turn important for many application scenarios. For instance, the determination of the particle size of plasmonic nanoparticles from their optical spectra is promising for fast characterization. Commonly used dynamic light scattering (DLS) measurements are influenced by morphological features (surface ligand, nanoparticle shape, roughness, *etc*.), deconvolution models for the correlation function and so on.[35] In contrast, determination of the particle size from the plasmonic response can be fast and straightforward[36], and has been studied in some previous works, especially for Au.[37][38] In order to obtain detailed information about a population of colloidal nanoparticles, the effect of all the different factors discussed above needs to be understood individually.

In this work, we first establish the effect of the particle size distribution explicitly by varying the standard deviation for a constant mean size while keeping the total number of atoms constant in a normal distribution. Apart from the plasmon band broadening and the expression of the multipolar modes, this also clarifies the effect of a higher volume contribution by larger nanoparticles for sizes larger than the mean. Then, a normal distribution was converted to a log-normal distribution to study the effect of the skewness in the distribution. Further, the effect of surface roughness and different optical constants from different literature are compared. The comparisons of theoretical results with experimental results was validated by reproducing the simple experiments of Bastús *et al*.[8] to synthesize Ag nanoparticles, and their optical spectra were recorded for comparison. It is clear that the mean is the most important parameter, while the standard deviation and skewness only exert a minor influence on the prediction of theoretical spectra with appropriate optical constants.

## 2. Theoretical formulation

### 2.1 Mie analytical solution for a distribution of nanoparticles

*Normal distribution*. The size distributions based on constant mass were obtained by a bottom-up approach with considerations to the colloidal reduction methods. Although a top-down synthesis method like pulsed laser ablation is conceptually completely different, the particle size distribution still follows similar trends owing to the universality of the central limit theorem.[39] For the calculations based on a fixed atomic population, the number of Ag atoms was fixed at 1.5055 x $10^{19}$ (based on 1 ml of 0.025 millimolar $AgNO_3$ solution) according to a generic one pot synthesis procedure reported by Bastús *et al*.[8] The reproducibility of the protocol was tested experimentally in order to ensure the validity of the theoretical assumptions and comparisons. Thus, this method emulates colloidal plasmonic nanoparticles of concentrations at the Beer Lambert limit in contrast to previous works on planer particle arrangement on a substrates.[40][41] The formation of polydisperse nanoparticles from the given number of metal atoms is described in the schematic in Figure 1(a). Regardless of the reducing agent and accompanying stabilizing reagents, the basic principle of the formation of the nanoparticles remains the same. For a fixed number of atoms, desired mean and standard deviation, the particle size distribution was obtained by iterative procedure as shown in the flow chart in Figure 1 (b). In the second step of this iterative process, the assumption for the number of nanoparticles has to be made judiciously towards obtaining a number of atoms as close to the initial fixed number as possible. For comparisons, the mean particle diameter was fixed and the standard deviation was varied which also obviously results in variation in the total number of nanoparticles. Assuming a normal distribution of the particle size, the Gaussian functions for the particle size distribution for a fixed number of particles is:

$$f(D) = \frac{N_{Total}}{\sqrt{2\pi}\sigma} exp\left[-\frac{1}{2}\left(\frac{D-\mu}{\sigma}\right)^2\right] \qquad (1)$$

The number of nanoparticles, $n$, in a predefined interval, $\Delta d$, can thus be obtained from eq. (1) as:

$$n = \Delta d \times f(D)$$

or

$$n = \Delta d \times \frac{N_{Total}}{\sqrt{2\pi}\sigma} exp\left[-\frac{1}{2}\left(\frac{D-\mu}{\sigma}\right)^2\right] \qquad (2)$$

In eq. (1) and (2), $N_{Total}$ is the number of total nanoparticles, $D$ is the mean diameter in the interval, $\mu$ is the mean diameter of all the nanoparticles and, $\sigma$ is the standard deviation. The particle size distribution for specific values of $\mu$ and $\sigma$ was obtained by the iterative procedure. The number of particles, $N_{Total}$ had to be first estimated with different values until the total number of atoms calculated from the entire distribution of nanoparticles with the given $\mu$ and $\sigma$ results in $1.5055 \times 10^{19}$. The molar mass and density values used in the calculations were 107.9 gm/mole and 10.5 gm/cm$^3$.

*Log normal distribution.* In order to study the effect of asymmetry in the distribution leading to skewness, the number of particles in a frequency interval, $\Delta d$ in a lognormal distribution can be given as:

$$n = \Delta d \times f(D) \qquad (3)$$

where, $f(D) = \frac{N_{Total}}{D\sqrt{2\pi}\sigma^*} exp\left[-\frac{1}{2}\left(\frac{\ln D - \mu^*}{\sigma}\right)^2\right]$ is the lognormal distribution function.

The location parameter $\mu^*$ and shape parameter $\sigma^*$ in eq. (3) are related to the mean and standard deviation of the distribution as:

$$\sigma^* = \sqrt{ln\left(\frac{\sigma^2}{e^{2ln\mu}} + 1\right)} \tag{4}$$

$$\mu^* = ln\,\mu - \frac{\sigma^{*2}}{2} \tag{5}$$

Thus, for a given mean diameter, $\mu$ and standard deviation, $\sigma$, defined for a standard normal distribution, a corresponding lognormal distribution can be obtained whose skewness can be given as:

$$Sk = (exp(\sigma^{*2}) + 2)\sqrt{exp(\sigma^{*2}) - 1} \tag{6}$$

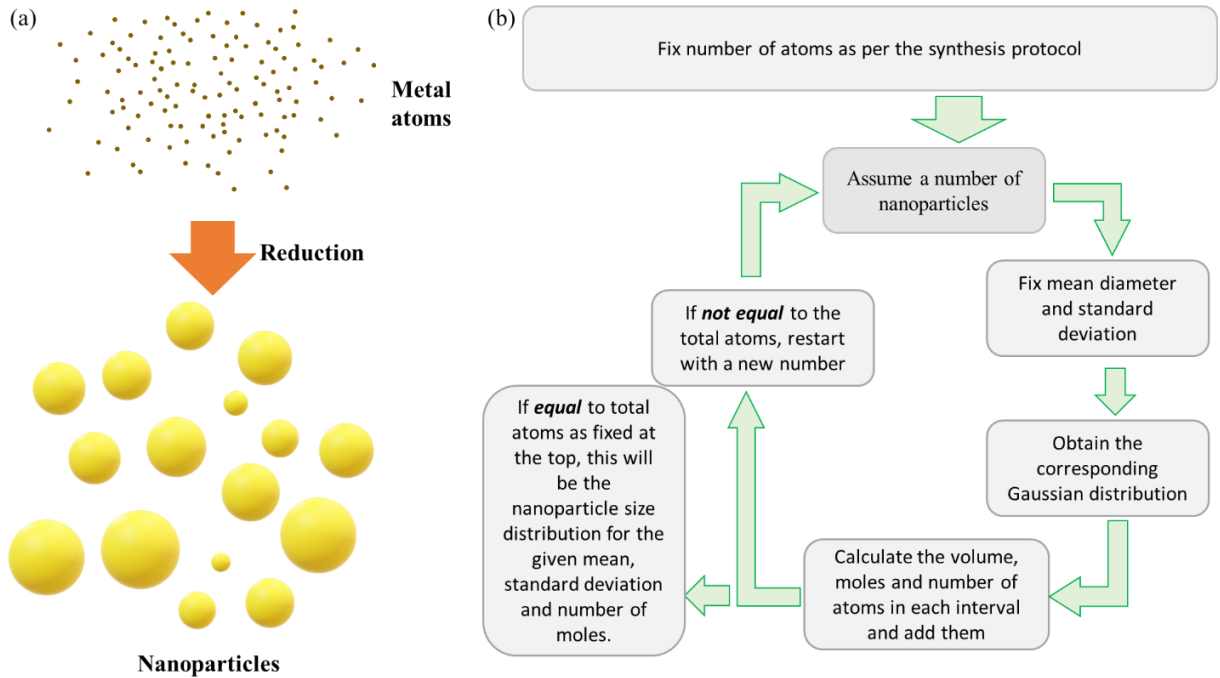

**Figure 1.** (a) Schematic describing the general colloidal synthesis route for plasmonic nanoparticles: formation of a distribution of nanoparticles from fixed number of atoms. (b) Flow chart describing how, for a fixed mole of atoms, distributions of different mean and standard deviation were obtained assuming a normal distribution.

*Mie theory calculations for a distribution of nanoparticles.* Upon obtaining the particle size distribution, the optical cross sections were calculated using a MATLAB script (supporting information). The underlying equations for the optical cross sections from the Mie analytical solution for spherical nanoparticles are well-known and are documented elsewhere.[36][42] For each set of wavelengths, corresponding complex optical constants of Ag and refractive index of the medium ($n$ = 1.33 for water), the extinction, absorption and scattering cross sections were computed for the entire particle population normally distributed about the mean size as shown in Figure S1. Specifically, the optical cross section computed for a given size (interval size: 0.1 nm) was multiplied with the number (frequency) of the nanoparticles in that interval/bin. Finally, the optical cross sections for each frequency interval were added in order to obtain the total optical cross sections for a specific wavelength. The important condition for the addition of the optical intensities of individual particles of the distribution is that the concentrations should be at the Beer-lambert limit so that the intensity is proportional to the concentration.

## 2.2 Calculation of optical response by numerical simulation

In order to investigate the effect of particle surface morphology, Maxwell's equations were solved for the scattered field in the frequency domain:

$$\nabla \times (\mu_r^{-1} \nabla \times E_{sc}) - k_o^2 (\varepsilon_r - j\frac{\sigma}{\omega \varepsilon_o}) E_{sc} = 0 \quad (7)$$

Where $\mu_r$, $\varepsilon_r$, and $\sigma$ are material properties namely relative permeability, relative permittivity and electrical conductivity respectively. Eq. (7) was solved numerically using a finite element method-based solver, COMSOL Multiphysics. Also, the roughness of the nanoparticle was artificially induced by meshing a spherical nanoparticle with coarse tetrahedral elements and then reconverting the mesh into a solid object. For the scattered field formulation, an isolated nanoparticle in a spherical computational domain with a perfectly matched layer (PML) was constructed. An incident plane wave was defined as the background electric field for which the scattered field solution was obtained.

## 2.3 Experimental section

For the verification of the literature data used herein, Ag nanoparticles were also synthesized experimentally following the protocol of Bastús *et al*.[8] All the chemicals were obtained from Sigma Aldrich. Briefly, 1 ml of 25 millimolar $AgNO_3$ solution was injected into a 100 ml boiling solution of sodium citrate (5 millimolar) and tannic acid under vigorous stirring. The concentration of tannic acid determines the mean size of the nanoparticles. Thus, for 15 nm and 36 nm nanoparticles, the tannic acid concentrations were 0.1 and 1 mM respectively. After the addition of $AgNO_3$, the solution was left to boil for another 10 minutes before rapid cooling in an ice bath. The nanoparticles were then centrifuged twice at 10000 rpm for 30 minutes for the removal of excessive citrate and tannic acid. The UV-Vis spectra of the nanoparticle colloids were recorded by a Shimadzu spectrophotometer (UV 2600).

## 3. Results and discussion

### 1. Effect of particle size distribution for fix number of Ag atoms

When the number of Ag atoms is fixed, for instance when starting from 1 mL of 25 mM AgNO$_3$ as in the experimental synthesis protocol by Bastús *et al.*[8], different scenarios for the particle size distribution can be envisaged. The mean size and the standard deviation are commonly reported as the primary indicative parameters of the distribution. Thus, naturally, a normal distribution is the first basic assumption for the size distribution. The deviations from the normal distribution will be discussed in the later sections. It is expected that with increasing standard deviation, *i.e.* widening of the size distribution, the collective plasmon band should also be widened as the contribution of the smaller and larger nanoparticles relative to the mean size will enhance the extinction intensities around the LSPR of the mean. In Figures 3 to 5, the effect of standard deviation in the normal distribution of nanoparticles with mean sizes of 20, 60 and 100 nm is delineated. It is important to note that a symmetric size distribution around a mean implies that the volume distribution will be skewed towards right due to bigger sizes. Since the optical absorption and scattering increase with increasing size, one may expect that with increasing standard deviation, the LSPR might red-shift due to a more dominant effect by the larger nanoparticles. However, it is clear that for 20 and 60 nm particles as the mean, the effect of standard deviation, in general, does not have a significant effect on the LSPR band position (Figure 2 and 3). Even a standard deviation of 33% of the mean did not result in any significant shift of the LSPR although the volume distribution is strongly shifted towards particles with larger diameters. While the band position remains constant, the effect of the standard deviation on the right-HWHM (half width at half maximum) is relatively stronger for the 60 nm average distributions than 20 nm average distributions (Figure 2 and 3). The LSPR position is determined by the condition at which the generalized multipolar resonance frequency, $\omega_l'(D)$, for a diameter $D$ and different eigenmodes $\{l = 1, 2…\}$, is equal to the incident wave frequency. Derkachova and Kolwas shows by numerically solving the dispersion relation for sodium that $\omega_l'(D)$ decreases first slowly (starting from radius: 0 nm) leading to a faster fall up to >200 nm, then to eventually become slow again.[43] This trend is particularly strong for the dipolar mode *i.e.*, $l = 1$ which is also the strongest in the extinction spectrum of smaller nanoparticles. Thus, it is expected that the effect of size distribution on the optical spectra will also exhibit an increasing trend with increasing mean size at least up to 150 nm. For the 60 nm mean, the quadrupolar mode ($l = 2$) is also quite visible as a shoulder on the left side of the spectra that gets smeared out with increasing polydispersity (Figure 3).

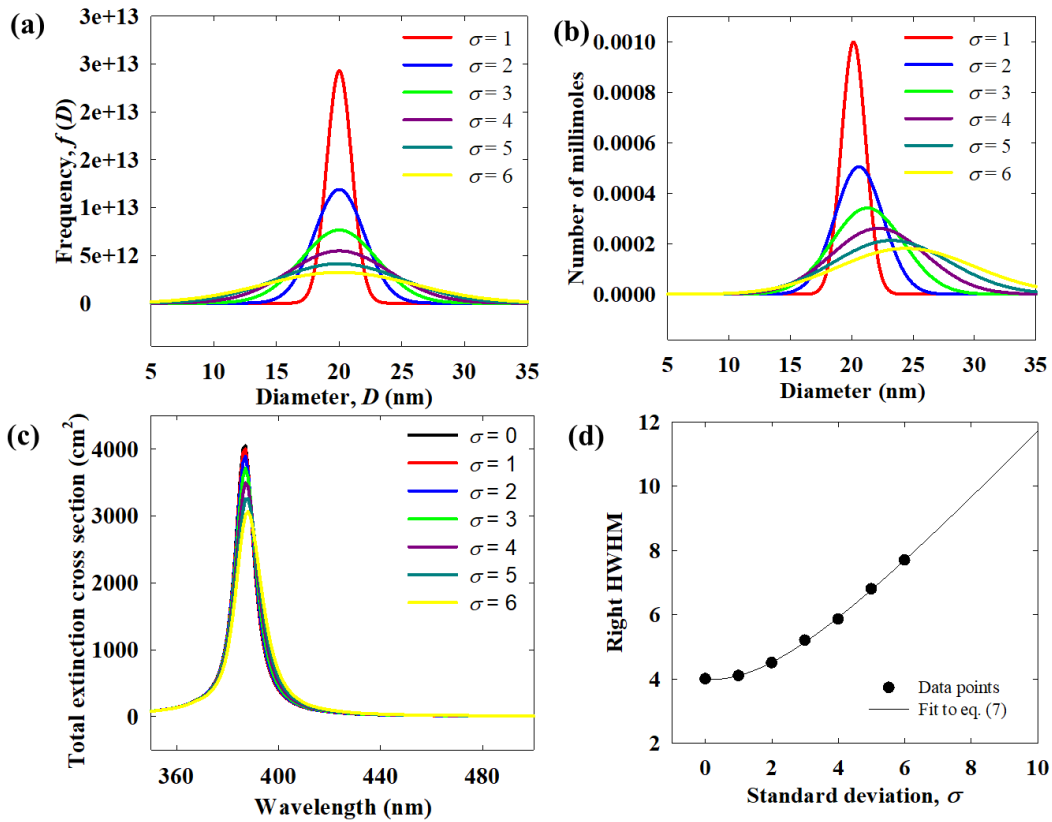

**Figure 2.** For a constant number of Ag atoms (0.025 mmol): (a) nanoparticle size distribution from eq. (1) for a fixed mean, $\mu$ = 20 nm and different values of standard deviation, $\sigma$ (b) corresponding atomic distribution with respect to size (c) total extinction cross section of the entire population of nanoparticles for $\mu$ = 20 nm and different values of $\sigma$ (in cm$^2$) (d) variation of the right-half width half maximum (HWHM) with increasing $\sigma$.

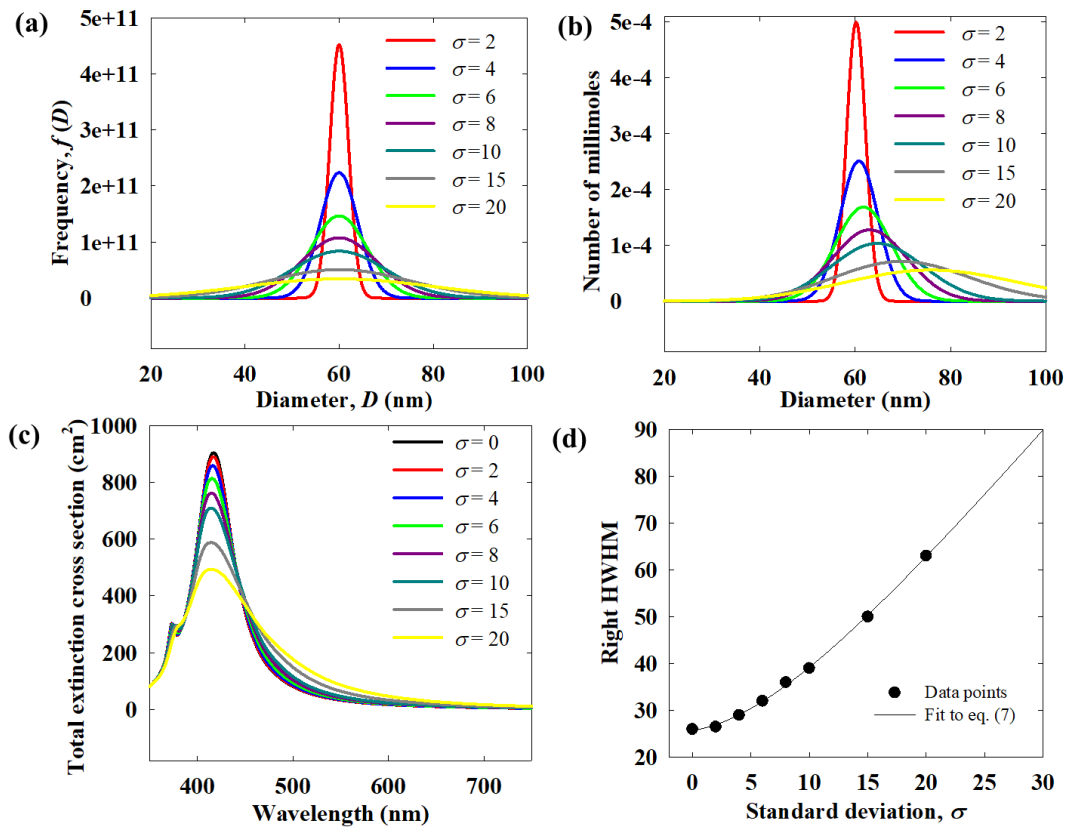

**Figure 3.** For a constant number of Ag atoms (0.025 mmol): (a) nanoparticle size distribution from eq. (1) for a fixed mean, $\mu = 60$ nm and different values of standard deviation, $\sigma$ (b) corresponding atomic distribution with respect to size (c) total extinction cross section of the entire population of nanoparticles for $\mu = 60$ nm and different values of $\sigma$ (in cm$^2$) (d) variation of the right half width half maximum (HWHM) with increasing $\sigma$.

For particles with a large mean diameter (for example, 100 nm, Figure 4), apart from the stronger size dependence of the LSPR shift in this size regime, the peaks due to both the dipolar (Eigenmode, $l = 1$) and the quadrupolar modes ($l = 2$) are strong.[44] Thus for the distributions with 100 nm as the mean, both these bands play a role in the overall spectra of the population. The effect of both of these aspects is clear from Figure 4, where the LSPR band is significantly influenced by the increasing standard deviation, thus resulting in weakening of both the modes to eventually become indistinguishable from one another. Importantly, similarities between these theoretical spectra and experimental ones also strongly depend on the correctness of the optical constants used in the theoretical simulations, which will be discussed in the later sections. It is convenient to correlate the dependence of the LSPR bandwidth (right HWHM) and the polydispersity, $\sigma$ for the prediction of the band-broadening for any standard deviation. An exponential growth function describes this relation empirically but with sufficient accuracy as given in eq. (8), where $Y_{HWHM}$ and $\sigma$ are right-HWHM and standard deviation:

$$Y_{HWHM} = c_o + c_1 x + c_2 \exp(c_3 \sigma + c_4) \tag{8}$$

In eq. (8), $c_o$, $c_1$, $c_2$ and $c_4$ are empirical coefficients. The statistical parameters in Table S1 show that the correlation fits the data points satisfactorily.

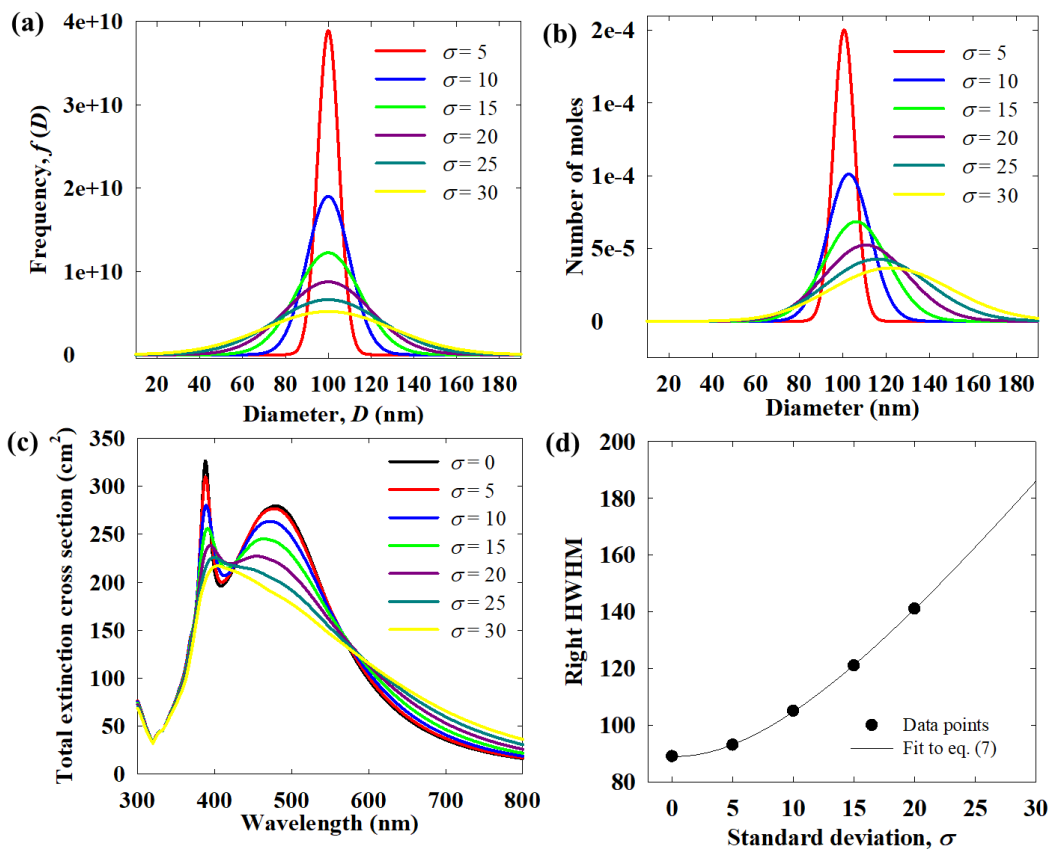

**Figure 4.** For a constant number of Ag atoms (0.025 millimole): (a) nanoparticle size distribution from eq. (1) for a fixed mean, $\mu = 100$ nm and different values of standard deviation, $\sigma$ (b) corresponding atomic distribution with respect to size (c) total extinction cross section of the entire population of nanoparticles for $\mu = 100$ nm and different values of $\sigma$ (in cm$^2$) (d) variation of the right-half width half maximum (HWHM) with increasing $\sigma$.

While the normal distribution is a simple approximation, it is known that in reality the formation of smaller clusters is actually favored, which leads to a skewness towards smaller sizes.[45][46] As per the coagulation model of Smoluchowski[47], this is a self-preserving size distribution during Brownian coagulation which takes place at the nucleation and crystallization stages of nanoparticle formation. In an alternative modeling approach, Kriss *et al.* showed that distribution of growth time determines the skewed lognormal distribution.[48] For smaller standard deviations, a normal and a lognormal distribution are indistinguishable. However, with increasing polydispersity, a lognormal distribution tends to be progressively skewed with respect to a normal distribution with the same mean and standard deviation. As shown in Figure 5 (a), for a mean of 60 nm, the distribution with a standard deviation of 20 nm is significantly skewed towards left with a skewness factor, $Sk$, of 1.037 in contrast to standard deviation of 10 ($Sk$ = 0.5045). It is, however, clear that even for highly polydisperse particles, the spectra for the lognormal distribution do not deviate much from those for the normal distribution. Even for a standard deviation of 33.3% of the mean, which can be described as a highly polydisperse population in view of the present literature on nanoparticle synthesis, the spectrum is only slightly blue shifted for the lognormal distribution. In case of the protocol reported by Bastús *et al.*, a normal distribution is thus a satisfactory assumption as the standard deviations are too low to have any significant effect. It is clear that in general under typical colloidal synthesis conditions, the mean size is the most important statistical parameter that determines the optical spectrum, while, polydispersity and skewness only have a minor influence.

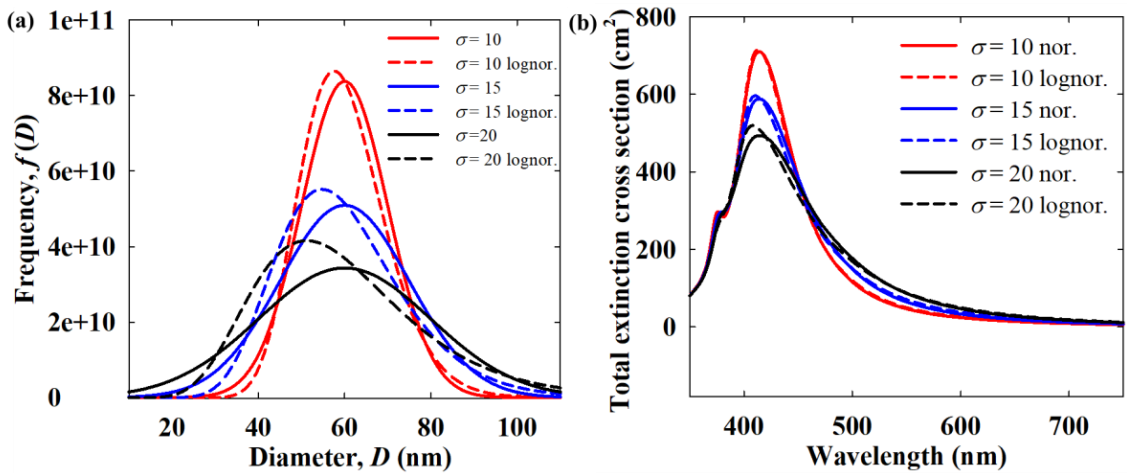

**Figure 5.** For a fixed NP mean of 60 nm: (a) comparison between normal distribution from eq. (1) and lognormal particle size distribution from eq. (3) for three different values of $\sigma$ (solid lines: normal distribution, dashed line: lognormal distribution) (b) corresponding total extinction spectra comparing normal and lognormal distributions. The skewness for the three values of $\sigma$ = 10, 15, 20 are 0.5045, 0.7656 and 1.037 respectively. (all calculations were done for a constant number of Ag atoms: 0.025 millimoles)

**3.2 The deviation of theoretical predictions from experimental optical spectra**

Another important aspect that determines the optical properties of plasmonic nanoparticles is the surface morphology, as the optical response is strongly influenced by the surface characteristics as well as shape.[49] González *et al*. demonstrated moderately pronounced effect of the surface morphology on the optical spectra for small Ag nanoparticles from 4 to 13 nm in size.[50] The morphological irregularities in the larger particles synthesized by Bastús *et al*.[8] is clear from their TEM images (Figure S2) and it is possible that they might be playing a significant role in shaping the optical spectra. In order to study the effect of such surface irregularities, the optical spectrum of an ideal spherical nanoparticle was compared to that of roughened nanoparticles, Figure 6. Although the artificially induced roughness resulted in a reduction of the total volume and irregularities on the surface, the LSPR position and the bandwidth do not change enough to drastically alter the spectrum. This is in line with the work of Linge *et al*., defining plasmon length for Au nanoparticles as the determining parameter of a plasmonic particle's optical spectrum, which is a characteristic path length of plasmon oscillation independent of the surface morphology.[51] Also, for larger sizes, the edge effects becomes progressively less important.

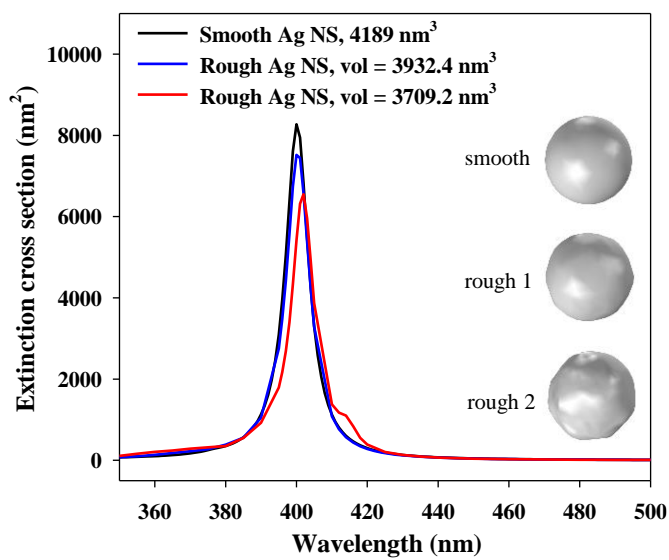

**Figure 6.** The effect of morphological irregularities on the optical spectra of Ag nanoparticles of diameter ~ 36 nm. (Optical constants used from Johnson and Christy[30])

The preceding discussion has shown that, in general, for a fixed mean, varying polydispersity of the population as well as moderate surface irregularities do not influence the optical spectra significantly. In view of the reference synthesis protocol for this study by Bastús *et al*.[8], the polydispersity, skewness and morphological irregularities are low enough to not be playing any significant role in spectral characteristics. Thus, the remaining primary factor deemed to be determining the spectral characteristics in theoretical simulations is the suitability of the optical constants used in the calculations. While the optical constants data from Johnson and Christy[30] work well for Au nanoparticles (Figure S3), there is a major mismatch for Ag nanoparticles, Figure 7 (a) and S4. In comparison, the constants from Rioux *et al*.[33] and Palik[29] have a much better performance in predicting the LSPR position. The complete spectra in Figure 7(b) show that, except from Palik[29], all the other optical data result in much narrower spectral bands as well as differences in the LSPR position. Since the recent data from Wu *et al*.[32] were obtained for monocrystalline Ag thin films, this difference is consistent with the existence of several crystalline domains even in colloidal Ag clusters of sizes as small as 5 nm.[52] Hence, the important factor here appears to be the degree of polycrystallinity or defects in the thin film samples that were used for the measurement of the optical constants, and the extent to which this matches the crystallinity of the colloidal Ag nanoparticles. From the comparisons done in this work, the optical constants from Palik[29] seem to be the best for the theoretical prediction of the optical spectra both in terms of LSPR position and plasmon bandwidth, Figure 7 (c) and (d). For the legitimization of the comparisons reported in Figure 7, the experiments of Bastús *et al*.[8] were successfully reproduced for this work and excellent agreement was obtained in the UV-Vis absorbance spectra for both the target average diameter of ~15 and ~36 nm. The standard deviation calculated out of a population of 500 nanoparticles remains within 15% of the diameter as they claim that the protocol yields highly monodisperse particles. Clearly, for 15 nm average size, the standard deviation has almost no effect on the spectra, while also for 36 nm average size, even a standard deviation much larger than their reported value does not lead to any significant qualitative/quantitative change, which is fully consistent with our findings above.

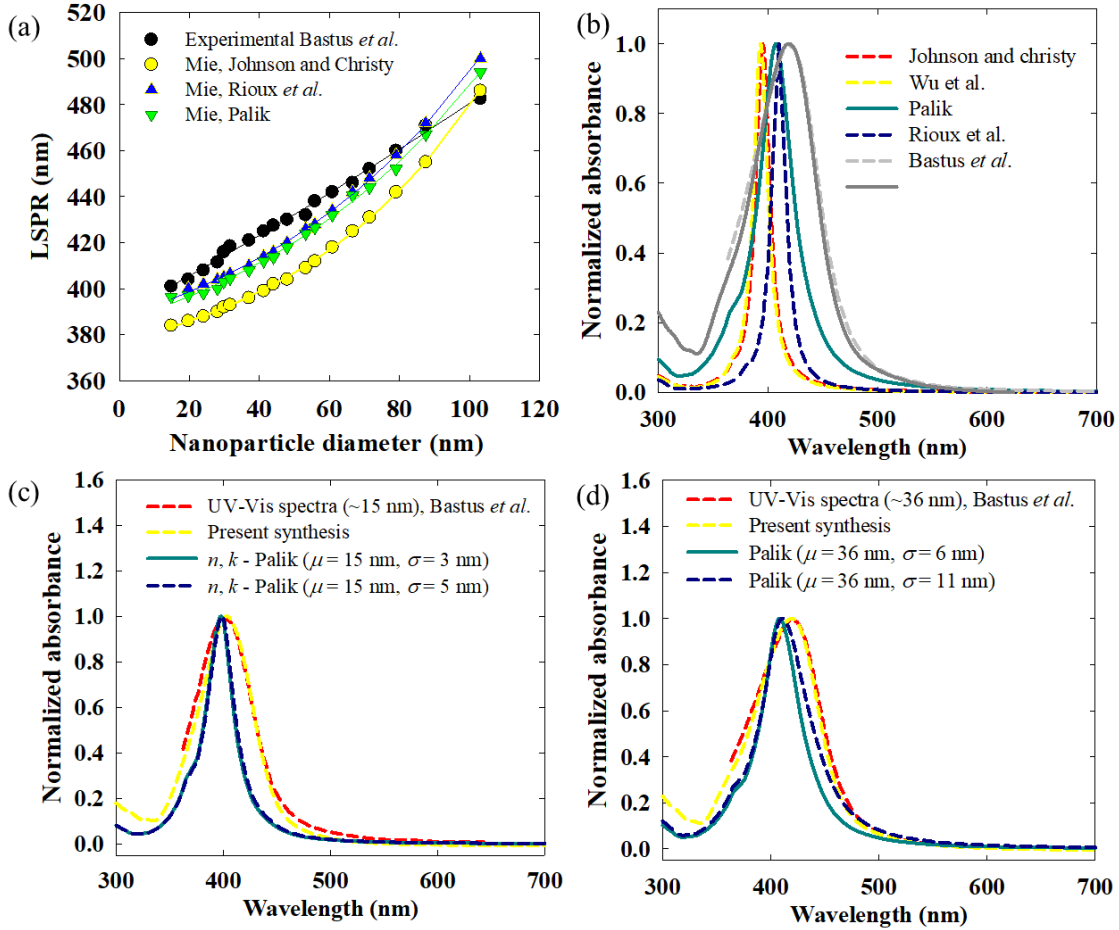

**Figure 7.** (a) Comparison between experimental (Bastús et al.[8]) and theoretical LSPR wavelength of single Ag nanoparticles of different sizes (only mean). The data is fitted to eq. (8) for the convenience in representation. (b) Comparison of different optical data from literature in terms of Mie spectra of a 36 nm nanoparticle. (c,d) Comparison of Bastús et al. data and present experimental spectra with theoretical spectra computed using Palik's optical data for 15 and 36 nm mean diameter.

In Figure 4 (c), for 100 nm mean distributions, the quadrupolar mode is quite strong in spectra calculated using the optical data of Johnson and Christy[30], especially for narrow distributions. It also smears the dipolar band in wider distributions, resulting in a plateauing between the quadrupolar and the dipolar peaks. However, in experimental spectra, such high intensities for the multipolar modes are not observed.[8] [53] In Figure 8, the expression of the dipolar and quadrupolar modes by the Palik's optical data in spectra obtained by applying Mie theory, is shown for different standard deviations and a constant mean diameter of 100 nm. While the quadrupolar mode is relatively weaker in this case, the experimental spectra still show a much weaker intensity resulting only in a faint shoulder. For wider distributions, the plateauing of the spectra due to equal intensities of the two modes is also quite strong. Nonetheless, the familiarity between the experimental spectra for 103.1 ± 7.4 particles and theoretical particles assert the overall suitability of Palik's data.

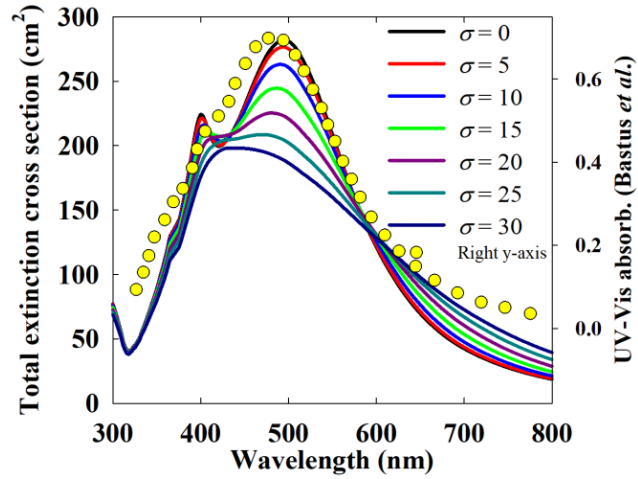

**Figure 8.** Performance of Palik's optical data in resolving the dipolar and quadrupolar modes of 100 nm mean distributions with varying polydispersity and comparison with 103.1 ± 7.4 nm Ag nanoparticles reported by Bastús *et al*.[8]

In order to correlate the dependence of the theoretical LSPR position on the size of Au nanoparticles, Haiss *et al*. implemented an exponential function, which despite being purely empirical, fits experimental data almost perfectly.[37] Since it is the mean that determines the LSPR position with minimal influence of the polydispersity and skewness, such a correlation allows one to predict the mean nanoparticle size directly from the optical spectra. The usefulness of such a correlation only obtained from theoretical results is, however, dependent on the suitability of the optical constants used. For Au, the theoretical predictions are quite satisfactory even with the commonly used Johnson and Christy's[30] optical data (Figure S3). A similar correlation can now also be defined for Ag nanoparticles where $c_o$, $c_1$ and $c_2$ are empirical constants:

$$\lambda_{LSPR} = c_o + c_1 \exp(c_2 D) \tag{9}$$

or

$$D = ln\,[(\lambda_{LSPR} - c_o)/c_1]/c_2 \tag{10}$$

In eq. (9) and (10), $\lambda_{LSPR}$ is the LSPR position and $D$ is the particle diameter. The quality of the fit for the different data sets given in Table 1 shows that eq. (9) correlates the wavelength dependence of LSPR quite well. Thus, eq. (10) can be used along with the appropriate empirical constants to predict the mean particle size simply based on the LSPR position. However, due to the larger differences between the theoretical and the experimental results for Ag nanoparticles (Figure 7), the correlations from theoretical Mie solution in

Table 1 can predict mean sizes (*D*) that are slightly off from the real (experimental) mean. In this regard, the correlations from Rioux *et al*. and Palik's optical data are better in terms of the closeness to the experimental data. One can also use the first correlation, which is based on the experimental data, directly rather than the theoretical size-LSPR dependence. In this context, it would be interesting to have a standard set of experimental data with complete characterization correlating the mean size to the LSPR position, to be used as a universally accepted reference for the practical implementation of the eq. (9) for the estimation of mean size from optical spectra.

**Table 1**. Fitting parameters with statistics correlating mean size and LSPR position in Figure 7 (a) with eq. (9).

|  | parameters | | | Statistics | |
| --- | --- | --- | --- | --- | --- |
| data | $c_o$ | $c_1$ | $c_2$ | $R^2$ | Adj. $R^2$ |
| Bastús *et al*. experimental | ~0 | 388.75 | 0.0021 | 0.9954 | 0.9948 |
| Mie (Johnson & Christy) | 355.6469 | 21.2390 | 0.0176 | 0.9996 | 0.9995 |
| Mie (Rioux *et al*.) | 354.0039 | 33.5452 | 0.0143 | 0.9993 | 0.9992 |
| Mie (Palik) | 349.5965 | 35.9518 | 0.0135 | 0.9976 | 0.9973 |

While the theoretical data in Figure 7 (a) follow a clear exponential trend over the reported size range, the experimental data tend to be more linear over this range although a very slow exponential growth also results in a decent fit. For comparison, the data from Agnihotri et *al*.[53] were inspected (Figure S5) and a similar size dependence was observed. Since the mean largely determines the LSPR position, this experimental trend is unlikely to be false due to any reason other than a wrongly estimated mean particle size. Given that both Bastús *et al*.[8] and Agnihotri *et al*.[53] characterized the size by TEM, these estimations depend on how representative the sample is with respect to the whole population. Based on the observed trends above, it appears that the experimental LSPR position-size relationship for Ag nanoparticles has a more linear tendency than that the theoretically predicted LSPR position-size relationship (Figure 7 (a)). Thus, the problem of theoretical prediction of optical spectra of Ag nanoparticles trickles down to finding the right optical constants. In this regard, Palik's data exhibit quite some differences from those from other sources[54], but also out-perform others in matching the experimental results. While the physical aspects down from the atomic level behind these observations is not in the scope of this study, the need of better experimental optical data suitable to the crystallinity of the colloidal nanoparticles remains.

### 3.3 Implications for dynamic Light Scattering (DLS)

The fact that for moderate polydispersity the assumption holds of a normal particle distribution, has interesting implications for particle size analysis with dynamic light scattering (DLS). Dynamic light scattering is a non-intrusive technique to estimate the particle size distribution. However, it is often difficult to estimate the true size distribution of nanoparticles with DLS as the hydrodynamic diameter strongly depends on the particle morphological features. Even for particles with a perfect spherical shape and smooth morphology, fundamentally, the intensity distribution given by DLS as the raw data for a fixed mean size shifts towards larger sizes with increasing polydispersity, Figure 9 (a) and (b). Thus, it is difficult to measure the true mean of the nanoparticle population as the polydispersity always leads to larger values than the real. For instance as reported by Bastús *et al.*[8], the mean from DLS for the 36.9 nm expected mean diameter (from TEM) population is 48.6 nm; a much higher value. Generally, the DLS mean is at least 25% above the expected TEM mean. Again, one cannot also rely completely on TEM data as during imaging one may tend to select the more monodisperse part of the population for better representation, while leaving out the extremes. This fundamental disadvantage of DLS measurements is besides the draw-back that the hydrodynamic diameter is estimated from the diffusivity, which in turn depends on the surface ligands.

Since it has been established that the normal distribution is a decent approximation for polydisperse particles obtained via colloidal synthesis methods and the mean can be approximated directly from the LSPR position, the standard deviation can be estimated from how much the intensity distribution peak obtained in DLS deviates from the mean due to the polydispersity. For instance, the intensity distribution peak in DLS for the 36.9 nm mean particles from Bastus *et al.*[8], lies at 43 nm (Figure S6Figure ). In contrast, the optical absorption spectrum suggests that the true mean diameter is ~ 34 nm, which is close to the reported mean of 36.9 nm from direct TEM analysis. The difference between the DLS intensity distribution peak and the true mean, 9 nm, is therefore indicative of the polydispersity. The larger this difference, the more polydisperse the sample. As shown in Figure 9 (b), the relationship between the population standard deviation and the theoretical intensity distribution peak can be correlated empirically with the following equation same as eq. (8):

$$Y_{int.\ dist.\ peak} = c_o + c_1\sigma + c_2 \exp(c_3\sigma + c_4) \tag{11}$$

In eq. (11), $Y_{int.\ dist.\ peak}$ and $\sigma$ are the scattering intensity distribution peak position and the standard deviation respectively, and $c_o$, $c_1$, $c_2$ and $c_4$ are empirical parameters. The goodness of the fit for the different data sets given in Table 2 shows that eq. (11) correlate the $\sigma$ dependence of $Y_{int.\ dist.\ peak}$ quite well. The relationship in Figure 9 (b) is an example of how for a fixed mean, the intensity distribution peak depends on the standard deviation correlated by eq. (11). Such a relationship can be derived for any mean size after the mean has been estimated from the LSPR position.

**Table 2**. Fitting parameters with statistics correlating intensity distribution peak, $Y_{int.\ dist.\ peak}$ and standard deviation, $\sigma$ (Figure 9 (b)) for distributions with 36 nm as the mean.

| Parameters | | | | | Statistics | |
|---|---|---|---|---|---|---|
| $c_o$ | $c_1$ | $c_2$ | $c_3$ | $c_4$ | $R^2$ | Adj. $R^2$ |
| 10.7299 | 3.0545 | 2.6707 | -0.119 | 2.2463 | 0.9997 | 0.9996 |

It is important to note that the accuracy of the direct conversion of the intensity distribution to a volume or population distribution is again subject to the accuracy of the optical constants used in the Mie theory calculations.[55] This is shown in Figure 9 (c), by using two different optical datasets yielding two different size distribution (FWHM). The higher frequency for Palik's[29] data is because of the fact that at 52 nm particle size (intensity distribution peak), the scattering cross section obtained using Palik's[29] data is 757 nm$^2$, higher than that obtained from Johnson and Christy[30], 620 nm$^2$. However, the position of the intensity distribution peak is not influenced by the quality of the data. Thus, the estimation of the standard deviation from Figure 9 (b) is unaffected by the choice of optical constants. Figure 9 (d) further corroborates the theoretical formulation of the size distribution by showing that the intensity distribution is not dependent on the wavelength of laser used in the DLS, as the laser type varies from manufacturer to manufacturer. Thus, the relationship between the intensity distribution peak and standard deviation shown in Figure 9 (b) is independent of the choice of optical constants and wavelength of illumination. It is however useful to use a wavelength closer to the LSPR, which is 530 in this case, so that the strength of the signal is higher.

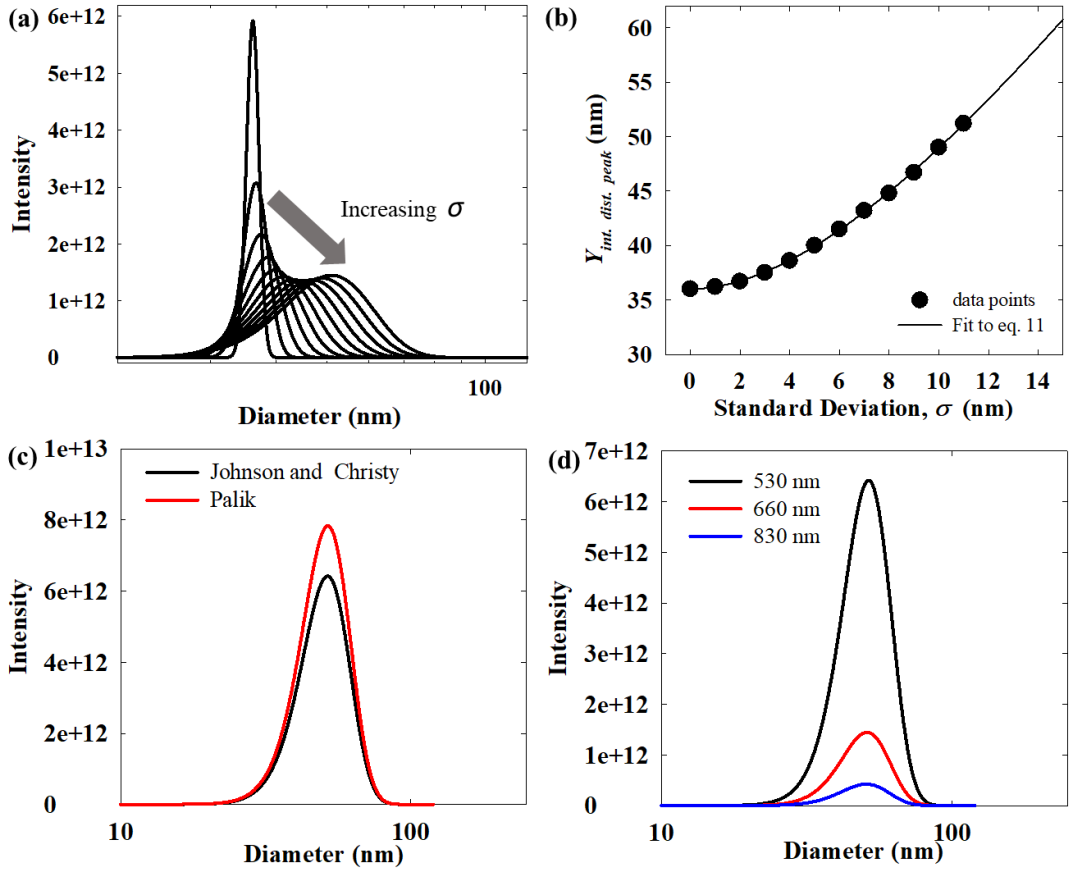

**Figure 9.** (a) Effect of polydispersity *i.e.* standard deviation, $\sigma$ on the theoretical scattering intensity distribution (theoretical raw DLS data) at 660 nm incident laser for a fixed mean of 36 nm. (b) Intensity distribution peak vs. $\sigma$ fitted to eq. (11). (c) The dependence of the theoretical scattering intensity distribution from optical data source. (d) The dependence of the theoretical scattering intensity distribution from incident laser wavelength.

## 4. Conclusion:

With the above results, the ambiguity around the role of particle size distribution, intrinsic optical properties and particle surface morphology in determining the collective optical response of colloidal plasmonic nanoparticles is resolved. The rather insignificant contributions of the polydispersity and the skewness for smaller colloidal Ag nanoparticles (~60 nm or less) synthesized via standard protocols in literature has been established. Generally, the mean size is the most important statistical parameter and determines the optical spectrum. The insignificance of the other statistical parameters implies that one only needs the suitable optical constants for theoretical calculations of the spectra. Also, it works well to assume a normal particle distribution as far as the optical spectra are concerned, since the tendency towards a lognormal distribution has no significant impact. It is possible that polycrystallinity plays a big role and Ag nanoparticles are mostly polycrystalline. We have shown that the position of the LSPR wavelength is not influenced by the particle size standard deviation when the mean is fixed, although the volume distribution shifts towards larger values as the standard deviation increases. For larger nanoparticles, however, the polydispersity and skewness become increasingly important as the size dependence of the LSPR increases with size. Also, the expression of the multipolar modes leads to broadening of the overall spectra. Based on the fact that the LSPR position is determined by the mean diameter, the mean diameter can indeed be accurately estimated from the optical spectra. For that, either carefully obtained standard experimental data can be used as universal reference, either theoretical calculations with suitable optical constants are required. Relying on the dependence of the DLS intensity distribution mean on the standard deviation, it is also possible to estimate the polydispersity from the difference between the mean size and the intensity distribution peak. While this study clears all the ambiguity regarding the primary factors determining the optical spectra of colloidal plasmonic nanoparticles.

**Acknowledgements**

R.B. and S.W.V. thank the University of Antwerp Special Research Fund for a DOCPRO4 doctoral scholarship.

**Graphical abstract**

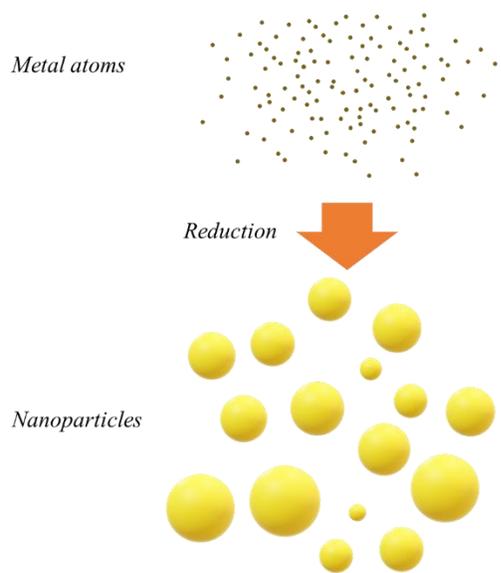
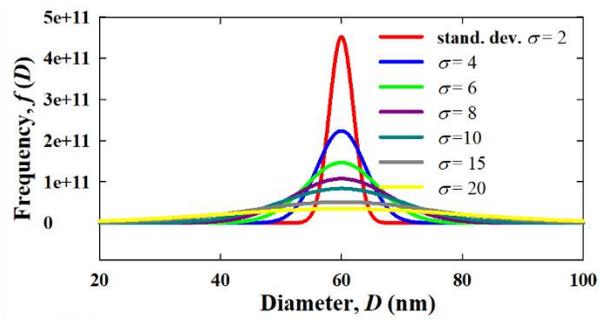
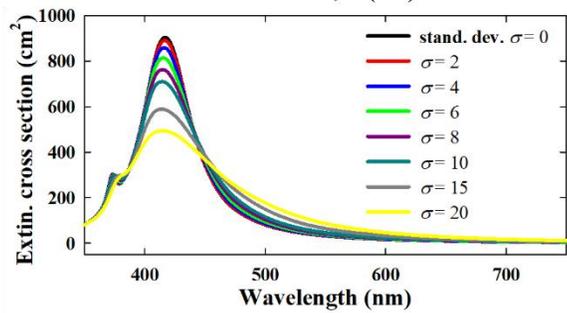